# Starch granules are instructive scaffolds for synergistic reinforcement and dissipation in hydrogel composites

Shirlaine Juliano[1], Jasmine Samaniego[2], Ian M Lillie[1], Geraldine Ramirez[1], Peter M Iovine[2*], Rae M Robertson-Anderson[1*]

[1]Department of Physics and Biophysics, University of San Diego, San Diego, CA 92110
[2]Department of Chemistry and Biochemistry, University of San Diego, San Diego, CA 92110

*randerson@sandiego.edu, piovine@sandiego.edu

## Abstract

A fundamental challenge in soft material design is the competition between rigidity and dynamicity, as stiffening mechanisms typically suppress energy dissipation. Here, we demonstrate that starch granules serve as instructive scaffolds that overcome this constraint, enabling the synergistic amplification of both elastic reinforcement and dynamic dissipation in hydrogels. We show that engineering the charge and structure of the filler-matrix interface enhances this synergistic response, which we propose arises from a dual-action physical mechanism: filler-induced polymer bundling of the polymer matrix provides structural reinforcement, while transient filler-matrix hydrogen bonding facilitates dissipation. Moreover, we reveal that binary blends of disparate filler species unexpectedly suppress these emergent properties, which we argue arises from enhanced entropic mixing. Our results provide a physical framework to overcome current design limitations in soft composites and sculpt their viscoelastic response from synergistic enhancement to strategic suppression for applications ranging from high-performance soft robotics to biomimetic tissue engineering.

**Introduction**

Composites pair two distinct materials, a matrix and reinforcing agent, in order to achieve synergistic properties not possible with a single component[1]. Soft matter composites are multiphase materials in which at least one constituent phase is mechanically soft—typically polymeric, elastomeric, or hydrogel-based[2]. These materials are used in diverse research areas including soft matter physics[3,4], biomaterials[5–7], polymer science[8,9], and food science[10]. Hydrogel matrices are especially attractive for biomaterials applications such as tissue engineering[11], wound healing[12], and biomimetics[13,14]; and both nano and micro-sized particles have been used as reinforcing agents in hydrogel composites[8,15].

Recent studies have suggested that starch granules may serve as powerful reinforcing agents in hydrogel composites[10,15–17]. For example, tissue-like properties have been demonstrated by embedding micron-sized native starch granules into polyacrylamide cross-linked networks[15]. An important conclusion from this work was that hydrogen bonding between the starch granules and the matrix plays an important role in energy dissipation. Upon stretching, the starch fillers strengthen the hydrogen bond network through granule-enabled strain focusing. In food science, the density and distribution of micron-sized starch granules in a glutenous network have substantial implications for regulating dough physical properties[18]. Finally, self-reinforcing starch films, in which the starch serves as both a matrix and an insoluble filler, have been developed[16,19] and shown to provide homogeneous phase structure and improved recyclability.

Starch granules are micron-sized naturally occurring polysaccharide particles composed of amylose and amylopectin. The balance of these two biopolymers determines much of the physical properties of granules, including gelatinization temperature, pasting properties, and swelling characteristics[20]. Both high amylose (HA) and high amylopectin (waxy) starches, like those used in this study, are available through selective breeding of starch-producing crops (e.g., corn). The size, shape, and molecular organization of the starch granule depends not only on the botanical source but also, in some cases, the history of chemical modification[21]. Chemical modification of granular starch is well established yet underexplored in soft material science. Under the appropriate conditions, a particular chemical modification can be carried out without disruption of the granule's three-dimensional structure, albeit at relatively low degree of substitution. This process equates to a type of solid-phase chemistry. Given the ability to modify starch granules with diverse chemical functionality, charges, and molecular recognition moieties, we suggest they represent a valuable and underappreciated class of microscale filler. For example, here we design and investigate a cationic starch derivative. The cationic moiety is a quaternary ammonium group and therefore insensitive to pH, and is present at low degrees of substitution (~0.03) to not disrupt the hierarchical granular structure.

We characterize the effect of concentration, charge and architecture of the starch granules on the bulk rheology and mesoscale organization of composites of gelatin and starch granules. We find that starch increases both energy dissipation and elastic storage of the gelatin, which we argue arises from transient hydrogen bonding between the starch and gelatin matrix. Moreover, altering the structure and charge of the granules, by increasing waxiness or adding positive charge, enhances these emergent effects, which we show arises from increased excluded volume and bundling of gelatin chains. At the same time, adding blends of any two types of starches to the gelatin matrix has a surprisingly weaker effect than any of the single starch additives, which we argue is a result of enhanced entropic mixing of the starches and gelatin that reduces the restructuring and disruption of the gelatin matrix responsible for the emergent rheological properties.

**Results**

We prepare composites of gelatin and starch granules and characterize the effect of starch concentration and surface chemistry on the bulk rheology and mesoscale structure of the composites. We find that composites with fixed gelatin concentration of 13.8 wt% and starch concentrations up to $[s] = 16$ wt% form 3D gel-like structures with the granules remaining largely homogenously dispersed throughout the gelatin matrix (**Fig 1A**).

**Starch granules enhance both the elastic plateau *and* the dissipative response of gelatin-based hydrogels.** To characterize the viscoelastic properties of the composites, we perform bulk linear oscillatory rheology measurements over 3 decades of frequencies. We measure the frequency-dependent storage and loss modulus, $G'(\omega)$ and $G''(\omega)$, which respectively quantify the elastic or solid-like response and liquid-like or viscous response of the material. In other words, elastic solids store energy while viscous fluids dissipate or 'lose' energy, so $G'(\omega)$ and $G''(\omega)$ are measures of how much a material stores or dissipates energy that is put into it by e.g., shear strain. We observe that adding 16% starch to the gelatin substantially increases both the elastic and viscous modulus (**Fig 1B**). However, the increase in $G''(\omega)$ is larger than for $G'(\omega)$ which we can clearly see by examining the loss tangent, $\tan\delta(\omega) = G''/G'$ which is >1 or <1 for viscous-dominated or elastic-dominated materials and is larger for systems with increased viscous dissipation (**Fig 1C**). The frequency-independent plateau in the elastic modulus, $G^0$ and $\tan\delta(\omega)$ being <1 over the entire frequency range for all starch concentrations confirms gel-like behavior in all cases, while the larger $G'(\omega)$ and $\tan\delta(\omega)$ values for starch composites suggests a complex contribution to the gelatin rheology which we explore further below.

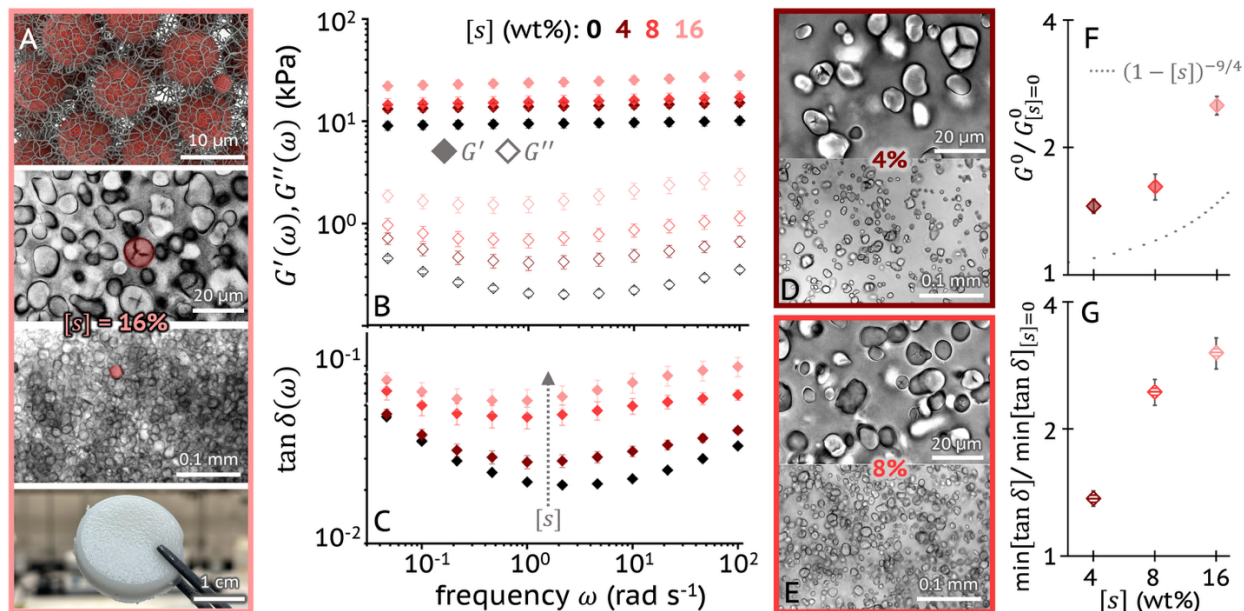

**Figure 1. Composite gels of starch granules and gelatin exhibit rich viscoelastic properties. (A)** Cartoon (top), bright field microscopy images (middle) and photo (bottom) of a composite comprising 16 wt% starch granules ([$s$] = 16%) and 13.8 wt% gelatin. Starch granules are shown in red in the cartoon and a single granule is highlighted in red in the microscopy images. **(B)** Linear viscoelastic moduli, $G'(\omega)$ (filled symbols) and $G''(\omega)$ (open symbols), and **(C)** loss tangent $\tan\delta(\omega)$ as a function of oscillation frequency $\omega$ for composites of gelatin and varying concentrations of starch granules: [$s$] (wt%) = 0 (black), 4 (maroon), 8 (red), 16 (pink). **(D,E)** Microscopy images of composites with [$s$] = 4% (D) and 8% (E) with the same magnification and imaging parameters as shown for [$s$] = 16% (A). **(F)** Plateau modulus $G^0$ and **(G)** minimum loss tangent, min[$\tan\delta$], each normalized by their value for pure gelatin ([$s$] = 0), and plotted as a function of [$s$]. Dashed grey line (F) is the predicted functional dependence for entangled polymers, displayed in the legend and described in the main text. Error bars in all plots denote standard error of the mean across all replicates and trials.

To determine how robust this emergent behavior is, we performed experiments on composites with 2× and 4× lower starch concentration ([$s$] = 8%, 4%). As shown in **Fig 1D,E**, both composites display similar largely homogeneous distribution of starch granules throughout the gelatin matrix. The impact on rheological properties is likewise similar to the 16% composite, though with a weaker increase in both $G'(\omega)$ and $G''(\omega)$ (**Fig 1B,C**).

To further examine the dependence of the rheological properties on [$s$], we evaluate the elastic plateau modulus $G^0$ (**Fig 1F**) and the minimum in $\tan\delta(\omega)$ (**Fig 1G**). The higher each value is the more that elastic-like and dissipative mechanisms, respectively, dictate the response. By plotting these metrics, normalized by their values for pure gelatin ([$s$] = 0), as a function of $[s]$, we find that both metrics are larger than their gelatin counterparts and steadily increase with $[s]$, with $G^0$ and min[tan $\delta$] reaching nearly 3-fold and 4-fold large values, respectively, as $[s]$ increases to 16%.

To understand this result, we first recall that for entangled polymers, the plateau modulus $G^0$ is predicted to scale with polymer concentration $c$ as $G^0 \sim (c/c^*)^{9/4}$, where $c^*$ is the polymer overlap concentration[22]. If we treat the starch granules as additives that are taking up volume in the composite and thus increasing the effective concentration $c$ of gelatin, we can estimate the relationship between $[s]$ and $c$ as $c \approx c_0/(1-[s])$ where $c_0$ is the original gelatin concentration. This estimate leads to the relation $G^0([s])/G^0_{[s]=0} \approx (1-[s])^{-9/4}$, plotted in **Fig 1F**, which reasonably captures the functional dependence on $[s]$ but systematically underestimates the magnitude of $G^0([s])/G^0_{[s]=0}$ by ~35-60% for all three concentrations. This result suggests that one of the primary contributors to the increased modulus is the effective concentrating of the gelatin matrix, but that another concentration-independent mechanism must also be at play to account for the universally larger values. We expand on possible additional mechanisms in the Discussion.

The substantial increase in the loss tangent is less intuitive, as an increase in the elastic response is typically coupled with a decrease in the dissipative response, rather than even more pronounced dissipation. One plausible mechanism for increased dissipation is by starch granules disrupting entanglements and interactions between gelatin chains, which are the primary contributors to the elastic response. If gelatin-gelatin interactions were traded for gelatin-starch interactions, and these interactions were more dissipative in nature, then we may expect an increased loss tangent. Previous studies have demonstrated the importance of hydrogen bonding in generating resilience and memory in starch-gelatin composites[15]. Transient hydrogen bonds between starch granules and the gelatin matrix allow the starch granules to move and rearrange in response to strains, thereby dissipating stress, while maintaining reinforcing connections with gelatin in the absence of strain. The result is a stronger material (higher modulus) that can more easily rearrange in response to strain to dissipate energy, thereby preventing rupture and fracture.

**Starch properties robustly tune the rheological response of starch-gelatin composites.**
To determine the extent to which starch-gelatin interactions and their impact on rheological properties can be tuned to, e.g., enhance emergent dissipation, we engineered composites comprising starches with different structure and charge. The data presented in Figure 1 is for high amylose (HA) starch granules comprising a mixture of predominantly linear amylose chains as well as branched amylopectin polymers. Here, we replace HA starch granules with 'waxy' ones that are almost entirely composed of amylopectin (**Fig 2A**), as well as HA starches with cationic-coated surfaces (**Fig 2B**). We hypothesize that the different architecture and charge will alter both the propensity for hydrogen bonding with the gelatin matrix, as well as the excluded volume that dictates the effective gelatin concentration, impacting $G''$ and $G'$, respectively.

We find that both modifications increase the effects observed with HA starches: increasing both moduli (**Fig 2C,D**) and the loss tangent (**Fig 2E,F**) more than the HA starches even at the lowest starch concentration ($[s] = 4\%$) (**Fig 2C,E**). The increase in $G^0$ is modest and most apparent for the cationic starches and for $[s] > 4\%$ (**Fig 2G**). However, the increase in the loss tangent is significant, with both modified starches increasing min[tan $\delta$ ] by a factor of ~2 above that of the HA starch at $[s] = 16\%$ (**Fig 2H),** which equates to a ~7-fold increase above that for pure gelatin. Interestingly, while the cationic starches lead to a more dramatic increase in dissipation compared to the waxy starches at the lowest concentration, for $[s] > 4\%$ their viscoelastic properties are indistinguishable from one another *and* markedly different than the HA case.

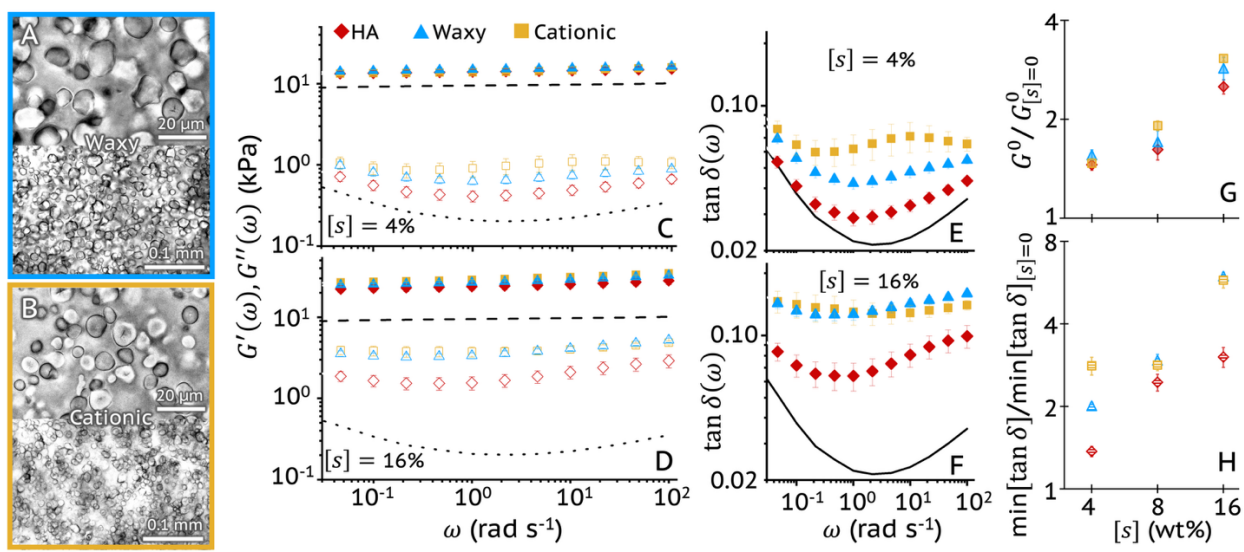

**Figure 2. Modifying starch structure and charge enhance dissipation in starch-gelatin composites. (A,B)** Brightfield microscopy images of composites with $[s] = 16\%$ Waxy (A, blue border) and Cationic (B, gold border) starches at high (top) and low (bottom) magnification. **(C,D)** Linear viscoelastic moduli, $G'(\omega)$ (filled symbols) and $G''(\omega)$ (open symbols), and **(E,F)** loss tangent tan $\delta(\omega)$ as a function of oscillation frequency $\omega$ for composites with HA (red diamonds), Waxy (blue triangles) and Cationic (gold squares) starches at $[s] = 4\%$ (C,E) and 16% (D,F). Black lines in C-F denote $G'(\omega)$ (dashed), $G''(\omega)$ (dotted) and tan $\delta(\omega)$ (solid) for pure gelatin ($[s] = 0$). **(G)** Plateau modulus $G^0$ and **(H)** minimum loss tangent, min[tan $\delta$], each normalized by their value for pure gelatin ($[s] = 0$) and plotted as a function of $[s]$ for composites with HA (red diamonds), Waxy (blue triangles) and Cationic (gold squares). Error bars in all plots denote standard error of the mean across all replicates and trials.

Comparing the images in Fig 2A,B with those of HA starches (**Fig 1A**), we see clear differences in size, shape and distribution on scales ranging from a few starches to dozens. Qualitatively, the waxy starches appear larger and more aspherical than the HA starches, in line with their known increased swelling properties. Cationic starches are similar in size to HA starches but appear to cluster together more than the others, possibly due to electrostatic repulsion with the weakly positive gelatin matrix (which is below its isoelectric point at the operating pH of 5.8) and rather low density of cationic sites on the starch granules. We expect these variations to play important roles in the rheological response as they alter both the surface area available for starch-gelatin interactions and the volume excluded from the gelatin matrix. Large scale clustering and larger particles may also more readily disrupt gelatin-gelatin interactions to promote dissipation.

**Blending distinct starches diminishes their impact on gelatin rheology.** Given the intriguing impact that starch granules have on the rheological properties of gelatin, and the important role of their specific properties in the degree to which the viscoelastic properties are modified, we next sought to determine if binary blends of the starches examined above could further enhance the emergent properties or offer an additional tuning knob for altering rheological properties. We prepared starch-gelatin composites with total starch concentrations of $[s] = 8\%$ and 16% with equal wt% of two different starches, e.g., ½$[s]$ each of HA and waxy (**Fig 3A**), waxy and cationic (**Fig 3B**), or HA and cationic (**Fig 3C**).

Compared to the different features seen in the microscopy images of single starch types (**Figs 1A, 2A,B**), there is far less distinction between the different binary blends at both high and low magnification (**Figs 3A-C, S3**). There are arguably more larger particles in the composites with waxy starches and more clustering with cationic starches, but the differences are more subtle than their single-type counterparts. Moreover, the starches appear to be more uniformly distributed throughout the field-of-view compared to single-type composites (**Figs S2,S3**), suggesting increased compatibility.

Turning to the rheological response, we discover that adding blends of starches has a surprisingly weaker impact on the viscoelastic properties of gelatin compared to adding a single starch species at the same concentration (**Fig 3D-I**). Specifically, all blended composites exhibit an increase in $G'$ and $G''$, similar to single-type composites, but the increase above that for pure gelatin is less (**Fig 3D-F**). Blends universally display less dissipation (lower $G''$ and $\tan\delta$) and reinforcement (lower $G'$) than that of their single-type counterparts. For example, $G'$, $G''$ and $\tan\delta$ and are larger for composites with either 16% waxy starches or 16% cationic starches compared to a 16% binary blend of waxy and cationic starches (**Fig 3E**). This result is rather counterintuitive, as one may expect that a binary blend would display rheological properties intermediate between the two single-type systems. Alternatively, composite systems often exhibit enhanced rheology and dynamics, meaning the response is greater than the sum of the responses of the two individual components[23–25]. Here we find quite the opposite effect. Blending two species results in a response that is weaker than that of either of the individual components.

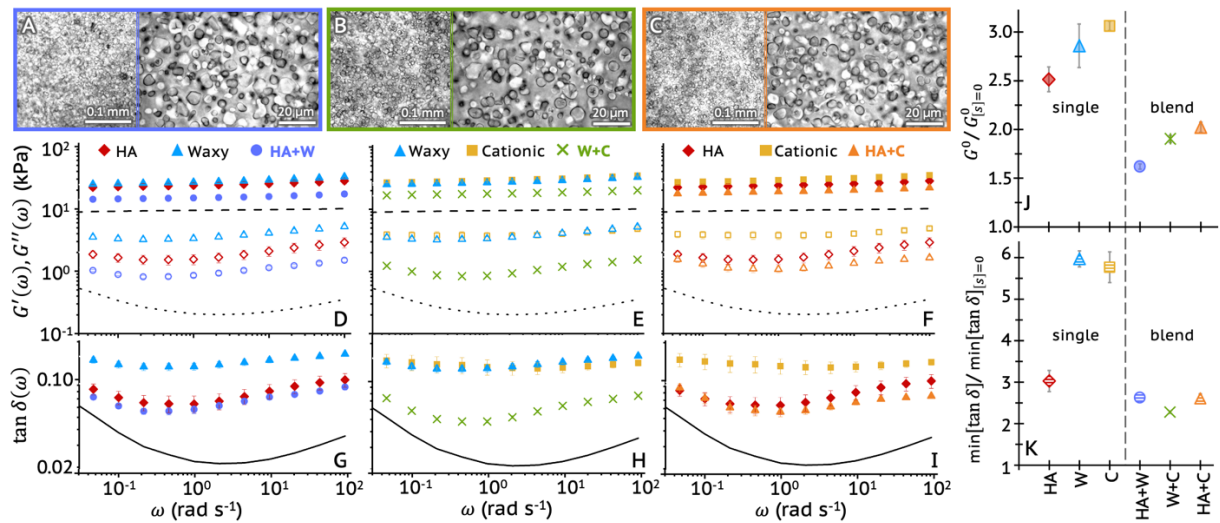

**Figure 3. Binary starch blends suppress the viscoelastic impact of single starch types on starch-gelatin composites. (A-C)** Brightfield microscopy images, at low (left) and high (right) magnification, of composites with total starch concentration $[s] = 16\%$ comprising 8% of each of two starch types: HA and Waxy (A, periwinkle border, H+W), Waxy and Cationic (B, green border, W+C), and HA and Cationic (C, orange border, H+C). **(D-F)** Linear viscoelastic moduli, $G'(\omega)$ (filled symbols) and $G''(\omega)$ (open symbols), and **(G-I)** loss tangent $\tan\delta(\omega)$ as a function of oscillation frequency $\omega$ for composites with $[s] = 16\%$ of binary blends shown in A-C and their single-type counterparts: (D,G) HA (H, red diamonds), Waxy (W, cyan triangles) and HA+W (periwinkle circles); (E,H) Waxy (W, cyan triangles), Cationic (C, gold squares) and W+C (green x's); (F,I) HA (H, red diamonds), Cationic (C, gold squares) and HA+C (orange triangles). Black lines in D-I denote $G'(\omega)$ (dashed), $G''(\omega)$ (dotted) and $\tan\delta(\omega)$ (solid) for pure gelatin ($[s] = 0$). **(J)** Plateau modulus $G^0$ and **(K)** minimum loss tangent, min[$\tan\delta$], each normalized by their value for pure gelatin ($[s] = 0$), and plotted for all data shown in D-I. The dashed vertical line separates the data for composites with single starch types and binary blends. Error bars in all plots denote standard error of the mean across all replicates and trials.

We can see this effect clearly by examining $G^0$ and min[$\tan\delta$] for 16% blends and single-starch composites (**Fig 3J,K**). Both metrics are universally larger for single-type composites compared to blends. Comparing the different blends, we find that $G^0$ is largest for those with cationic starches, similar to the enhanced reinforcement offered by cationic starches in single-type composites (**Fig 3J**). Moreover, while blends exhibit only modestly decreased dissipation (i.e., lower min[tan $\delta$]) compared to HA starch composites, we find that min[tan $\delta$] is >2-fold lower for blends compared to waxy and cationic composites (**Fig 3K**).

**Discussion and Conclusion**

To understand the intriguing rheological properties afforded by introducing starch granules into gelatin matrices, and the impact on granule properties and mixing, we first summarize our results by plotting the normalized values of min[tan $\delta$] versus $G^0$ for all composites: single starch types at $[s] = 4\%$, 8%,16% and binary blends at $[s] = 8\%,16\%$ (**Fig 4A**). The data confirm that starch fillers universally increase the dissipation and elastic storage of gelatin hydrogels, shown by all data points in Fig 4A lying above min[tan $\delta$]/min[tan $\delta$]$_{[s]=0}$ = 1 and $G^0/G^0_{[s]=0} = 1$.

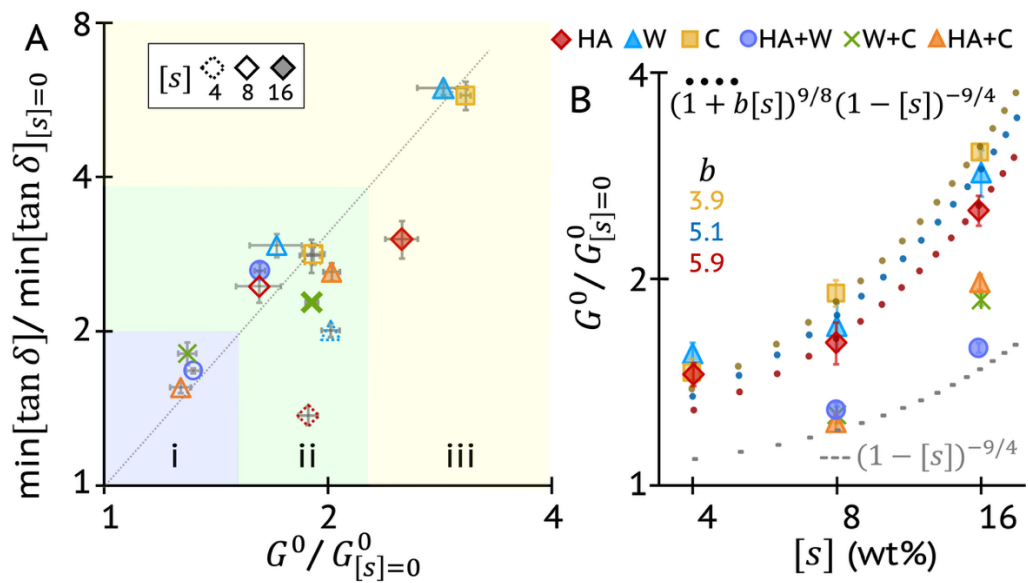

**Figure 4. Composites of starch granules and gelatin display a wide range of nontrivial rheological properties tuned by the starch properties and concentrations.** (**A**) Phase map of linear viscoelastic properties, quantified by the paired values of the normalized loss tangent and plateau modulus ($G^0/G^0_{[s]=0}$, min[tan $\delta$]/min[tan $\delta$]$_{[s]=0}$), for composites with all starch concentrations [$[s] = 4\%$ (open dotted line symbols), 8% (open symbols), and 16% (filled symbols)] and types [HA (red diamonds), Waxy (blue triangles), Cationic (gold squares), HA+Waxy (periwinkle circles), Waxy+Cationic (green x's) and HA+Cationic (orange triangles)]. The data naturally clusters into three phases marked by the strength of the coupled increase in dissipation (tan $\delta$) and reinforcement ($G^0$): (i, purple shading) weak, (ii, green shading) ample, and (iii, yellow shading) strong. The dashed line is a guide to the eye to show the coupling between $G^0$ and min[tan $\delta$]. (**B**) $G^0/G^0_{[s]=0}$ versus $[s]$ for single-type and blended starch composites with fits of the single-type data (color-matched dotted lines) to the function shown that incorporates bundling. The degree of bundling, quantified by the parameter $b$ determined from the fits, is listed for each starch type in color-matched font. Grey dashed line shows the predicted $[s]$ dependence without bundling (function shown in grey font), which the blended composite data follows.

Interestingly, for nearly all composites, we also observe that increases in min[tan $\delta$] are directly correlated with increases in $G^0$ and that increases in min[tan $\delta$] are larger than those for $G^0$, as seen by the dashed guide line (**Fig 4A**). This result is rather nontrivial given that increases in dissipation are typically coupled with decreases in elastic response, as a material trades elastic storage mechanisms for viscous dissipative ones. Moreover, comparing all composites, we find that the data naturally clusters into three phases with differing degrees of rheological changes: (i) weak ($[s] = 8\%$ blends), (ii) ample ($[s] = 16\%$ blends, $[s] = 4\%$ and 8% single-types), and (iii) strong ($[s] = 16\%$ single-types).

As suggested by ref [15], the increased dissipation is likely due to transient hydrogen bonding between starch granules and gelatin. These weak bonds can be easily broken during straining to allow starches to move and rearrange to dissipate stress. The increase in min[tan $\delta$] with starch concentration aligns with this interpretation as there would be more sites available for hydrogen bonding.

The higher dissipation for the cationic and waxy composites and lower dissipation for blends, suggests that larger starches and starch aggregates allow for more hydrogen bonding with the gelatin matrix. The physical picture is that larger starch entities can more effectively disrupt elastic-like gelatin-gelatin interactions, which are traded for more dissipative gelatin-starch interactions. Conversely, smaller, more dispersed starch particles may be able to be effectively embedded within the matrix without substantial disruption to gelatin-gelatin interactions that would increase the dissipation. Moreover, larger granules and aggregates with more aspherical configurations and surface roughness likely lead to more interfacial friction with the polymers that results in enhanced dissipation. Likewise, this increased friction renders the motion of the large starch constructs more energetically costly (dissipative).

To understand how the varied mesoscopic structures and gelatin-starch interactions described above, that underlie the increased dissipation, may also lead to correlated increases in elastic reinforcement, we revisit the excluded volume argument from the previous section (**Fig 1F**). Namely, starch granules take up volume in the composite that the gelatin can no longer occupy which effectively concentrates the gelatin above its nominal concentration. This mechanism roughly captures the concentration dependence but underestimates the magnitude of $G^0$ for all single-type composites, with the waxy and cationic starches displaying larger deviation from the prediction than HA starches (**Fig 4B**).

This larger deviation for waxy and cationic starches may suggest an effectively larger volume fraction excluded to the gelatin compared to HA starches at the same mass density. For waxy starches, this effect is due to the higher propensity to swell, which is evident in Fig 2A; while for cationic starches, the positively charged surfaces provide a greater repulsion with the gelatin, which is below its isoelectric point and thus slightly positive, rendering the effective volume excluded to the gelatin larger. This electrostatic repulsion also likely underlies the clustering of the cationic starches (**Fig 2B, S2**). The reduced volume leads to a larger effective gelatin concentration and thus a larger elastic modulus via the expression, $G^0 \sim (c/c^*)^{9/4}$.

Interestingly, the mixed composites follow the excluded volume prediction much more closely than the single-type composites (**Fig 4B**); and the granules appear to be more homogenously distributed throughout the composite (**Fig 3A-C**). Taken together, it appears that excluded volume effects contribute to the increase in $G^0$ for all composites, but that additional mechanisms are also at play for those with more clustering and/or mesoscopic heterogeneity, leading to an increase in $G^0$ above the excluded volume prediction.

To quantitatively verify the architectural differences between the composites described above, we compute the spatial image autocorrelation function $g(r)$ of the brightfield images of the composites to assess the average size and separation between starch granules (**Fig 5**). Evaluating the lengthscale over which pixels remain correlated with one another provides a measure of the average granule radius $\xi_g$, which we define as $g(\xi_g) = \mathrm{e}^{-1}$ based on the roughly exponential decay of $g(r)$ at short distances (**Fig 5Ai,Bi**). Following an initial decay to a local minimum, each $g(r)$ curve reaches a local maximum at a distance $\xi_s$, which is indicative of the average spacing between particles.

We find that $\xi_g$ is larger for composites with waxy starches compared to HA and cationic ones, confirming that waxy granules swell to larger sizes than their HA counterparts. We also find that $\xi_s$ is largest for cationic starches, in line with the increased aggregation that would create large starch-void regions between starch-rich clusters. Likewise, the larger $\xi_s$ for waxy compared to HA starches indicates enhanced clustering that is evident in Fig 2A. Examining the blended composites, we find that all $g(r)$ curves decay more rapidly with distance than their single-type counterparts (**Fig 5B**). This faster decay is quantified by the generally smaller particle size $\xi_g$ and separation $\xi_s$ for blends compared to single-types (**Fig 5Bii**), with average values of $\xi_g \simeq 2.2$ µm and $\xi_s \simeq 12$ µm versus $\xi_g \simeq 2.6$ µm and $\xi_s \simeq 20$ µm. This substantially smaller separation distance for blended composites is a key indicator of increased compatibility and reduced clustering, which appears to dampen both dissipative and reinforcing starch-gelatin interactions such that excluded volume arguments can more readily describe their behavior.

To understand possible mechanisms that drive the increased $G^0$ for single-type starches above the excluded volume prediction, we recall that a common phenomenon in polymer composites is bundling of polymer chains due to entropic depletion interactions[26–28]. Rather than forming an isotropic network, as polymers are forced into closer proximity they can nematically align to form bundles. The network then becomes a network of bundles which are stiffer, i.e., have a larger persistence length $p$, and thus larger radius of gyration $R_G$, via the relation $R_G \sim p^{1/2}$ [29].

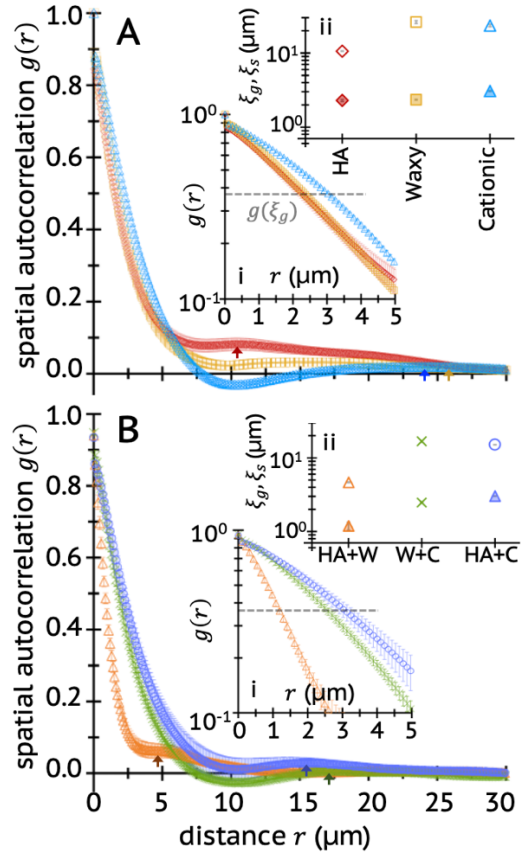

**Figure 5. Starch architecture, charge and blending dictate starch size and spacing between starch granules.** Spatial image autocorrelation function $g(r)$ versus distance $r$ for composites with a **(A)** single starch type or **(B)** binary blend. Arrows indicate $g(\xi_s)$ for the data with the matching color. Insets: (i) Zoom-in of $g(r)$ with dashed horizontal line denoting $g(\xi_g) = e^{-1}$; the distance at which the line intersects the data is the effective particle radius $\xi_g$. (ii) Small and large correlation lengthscales, measures of the average size, $\xi_g$ (filled symbols), and separation, $\xi_s$ (open symbols), of starch granules for each composite type, labeled on the x-axis. The colors and symbols apply to all panels.

If each polymeric entity (i.e, single polymer of molecular weight $M$ or bundle of $n$ polymers with molecular weight $nM$) is larger, then the overlap concentration is lower, according to the expression $c^* \approx (3M/4\pi N_A)R_G^{-3}$, making $c/c^*$ larger. Specifically, comparing a network of bundles each comprising $n$ polymers and having a persistence length $\sim np$ with that of single polymers with persistence length $p$ yields

$$\frac{c_n^*}{c_1^*} \approx \left(\frac{nM}{M}\right)\left(\frac{R_G}{R_{G,n}}\right)^3 \sim n\left(\frac{p}{np}\right)^{3/2} \sim n^{-1/2}.$$

Combining this relation with the concentration dependence of $G^0$ provides the expression $G_n^0 \sim (n^{9/8})(c/c_1^*)^{9/4}$ and the ratio $G^0([s])/G_{[s]=0}^0 \approx (n^{9/8})(1-[s])^{-9/4}$. For $[s] = 16\%$, the predicted ratio without considering bundling is $G^0([s])/G_{[s]=0}^0 \approx (1-[s])^{-9/4} \approx 1.5$ while the measured value is ~2.6. A bundling number of $n \approx 3.9$ would yield the experimentally measured increase from the pure gelatin case. We also expect that $n$ decreases with decreasing concentration with the limit of $n = 1$ for $[s] = 0$. If we assume a linear relationship, $n[s] = 1 + b[s]$, then the ratio becomes

$$G^0([s])/G_{[s]=0}^0 \approx (1+b[s])^{9/8}(1-[s])^{-9/4}.$$

Fitting the HA starch data to this model provides $b \approx 3.9$. Applying the same procedure for waxy and cationic starches yields $b \approx 5.1$ and ~5.9, respectively. These values align qualitatively with the physical picture that cationic starches are most effective at bundling gelatin, followed by waxy starches, while HA starches are the weakest bundling agents.

Within this model, the highly suppressed impact of mixed starches on the rheology suggests that hydrogen bonding and gelatin bundling are suppressed. Moreover, this suppression is expected to arise due to the reduced clustering and increased observable miscibility of the starches and gelatin (**Fig 5**).

The question remains as to why the mixed starches more readily mix with the gelatin compared to single starch types. Here, we consider the entropic gain associated with mixing multiple independent species: $\Delta S \sim - \sum x_i \ln x_i$ where $x_i$ is the molar fraction of species $i$ and the sum is over all species. From this expression, we can see that the more components in a system the greater entropic gain upon mixing due to the larger increase in volume available to each species. Specifically, if we assume $x_i$ is proportional to $[s]$, and the entropic gain to the gelatin $\Delta S_g$ is the same for any composite of a given $[s]$, then we can express the entropic gain from mixing a composite comprising a concentration $[s]$ of a single type of starch ($\Delta S_1$) or a concentration ½$[s]$ of each of two starch types ($\Delta S_2$) as:

$$\Delta S_1 \sim \Delta S_g - [s]\ln[s],$$

$$\Delta S_2 \sim \Delta S_g - 2\left[\frac{1}{2}s\right]\ln\left[\frac{1}{2}s\right] \sim \Delta S_1 - [s]\ln\left[\frac{1}{2}\right].$$

The extra term in $\Delta S_2$ drives a system with two starch types to mix more readily than a composite with only one. The increased miscibility, in turn, reduces the effect on the rheological properties. Notably, this result also highlights the distinct properties of the different starches. If they were indistinguishable then there would be no increased entropic drive to mix with one another.

In conclusion, we have established a physical framework for the simultaneous amplification of structural reinforcement and energy dissipation in bio-derived hydrogel composites. Our findings reveal that starch granules are not merely passive additives but instructive structural templates that bypass the traditional trade-off between dissipation and reinforcement in soft composites. We show that the efficacy of the synergistic amplification of dissipation and reinforcement afforded by starch granules can be strategically modulated by granule charge and structure. We argue that this synergy is driven by a dual-action mechanism: the formation of reinforced polymer bundles that densify the elastic network and interfacial hydrogen bonds that facilitate dynamic dissipation. Moreover, we demonstrate that increased entropic mixing in composites with binary blends of starches effectively suppresses the emergent properties, providing a counterintuitive 'off-switch' for material modulation. By demonstrating a sustainable pathway to simultaneously enhance or suppress these coupled mechanical responses, our work provides a predictive toolkit for the next generation of high-performance soft matter, with broad implications ranging from biomimetic tissue scaffolds and tunable food textures to energy-dissipating impact materials and soft robotics.

**Methods**

*Starches*. Three different granular starches were used in this study: high amylose corn-derived starch (HA), high amylopectin corn-derived starch (waxy), and cationic high amylose corn-derived starch (cationic). The cationic high amylose corn-derived starch (cationic) was a trimethyl ammonium chloride starch ether derivative derived from traditional cationization using either 2,3-epoxypropyltrimethylammonium chloride or 3-chloro-2-hydroxypropyltrimethylammonium in aqueous sodium hydroxide[30,31]. The degree of substitution (DS) for the cationic starch was approximately 0.03 as determined by $^1$H NMR in $d_6$-DMSO. All starches were stored at room temperature in sealed containers prior to their incorporation into the composite hydrogels.

*Gelatin-Starch Composites*. Composite gels of starches and gelatin comprised porcine gelatin, Type A, 300g Bloom (Millipore Sigma) and either a single type of starch (HA, Waxy, Cationic) or a blend of two of the three starch types. We defined starch concentration $[s]$ (wt%) as the weight percent relative to the total sample mass and fixed the total mass of the starch-water mixture at 7.5 g. We prepared composites at starch concentrations of $[s]$ = 0, 4, 8, and 16 wt% by, respectively, combining 0, 0.3, 0.6 and 0.9 g of starch and 7.5, 7.2, 6.9 and 6.6 mL of water. In all cases, we fixed the gelatin mass at 1.2 g (13.8 wt%). For the composites comprising binary blends of starches, we divided the total starch mass at a given wt % equally between the two starch types, while maintaining everything else constant.

*Gel Preparation.* To prepare the composites, we added the appropriate mass of starch to a Teflon beaker with an inner diameter of 3.6 cm and a depth of 2 cm followed by the corresponding volume of deionized water. We heated and continuously stirred the starch-water for 1 hour, maintaining the sample at 45 °C. We then slowly added 1.2 g of porcine gelatin powder and continued to stir until the gelatin was fully dissolved. To induce gelation, we incubated the composite solution at 4 °C for 15 minutes. The result is a disk with diameter ~2 cm and height ~0.5 cm. We transferred the gel to a humidity-controlled chamber which we sealed and kept at 4 °C overnight before performing experiments.

*Rheology.* We performed bulk linear rheology using a Hybrid Rheometer 3 (DHR-3, TA Instruments) with a 40 mm stainless steel parallel-plate geometry. We loaded each sample onto the lower plate, which was maintained at 22 °C using a Peltier temperature control system, and lowered the upper plate to a fixed gap height of 6.3 mm. We equilibrated samples for 3 minutes before collecting data.

We characterized the linear viscoelasticity (LVE) of the composites using small-amplitude oscillatory shear measurements. We performed strain sweep measurements to identify the linear viscoelastic regime (**Fig S1**), from which we determine a strain of 5% for frequency sweeps. We performed linear oscillatory measurements at 11 frequencies logarithmically spaced between $\omega = 0.04$ rad s$^{-1}$ and $\omega = 100$ rad s$^{-1}$ from which we determined the frequency-dependent storage modulus $G'(\omega)$ and loss modulus $G''(\omega)$. We also computed the loss tangent as a function of frequency via the relation $\tan\delta(\omega) = G''(\omega)/G'(\omega)$. For each composite, we performed experiments on three independent replicates and performed two consecutive sweeps for each replicate. The frequency-dependent data and error bars presented in the manuscript are the averages and standard error of the mean (SEM) across all replicates and sweeps. For each trial, we computed the elastic plateau modulus $G^0$ by averaging across all $G'(\omega)$ values and identified the minimum value of $\tan\delta(\omega)$. Average $G^0$ and min[$\tan\delta$] data presented in the manuscript are averages and SEM of the values determined from the individual trials.

*Microscopy.* To image the starch granules within the composites, we performed brightfield microscopy on thin sections of the prepared composites prior to rheology measurements. For each sample, we shaved a ~100 μm thick slice from the top surface of the gel disk using a razor blade. We placed the slice on a glass microscope slide, surrounded the periphery with vacuum grease and placed a No. 1 glass coverslip on top. The resulting airtight chamber kept the slice secured in place and prevented evaporation during imaging.

We imaged all samples using an Olympus IX73 inverted microscope with a 100 W halogen lamp (Olympus U-LH100L) and either a 60× 1.42 NA oil immersion objective (Olympus) or 20× 0.5 NA air immersion objective (Olympus). For each sample and magnification, we captured 5-10 images, 1920×1440 square-pixels in size, at different $x, y$ locations, using an ORCA-Fusion BT Digital CMOS (Hamamatsu) camera and HCImage (Hamamatsu) software. For each composite we performed imaging experiments on two independent replicates resulting in 10-20 images for each magnification (see **Figs 1A,D, 2A,B, 3A-C, S2, S3** for example images).

To quantify composite structure, we performed spatial image autocorrelation (SIA) analysis on the high magnification images using custom python scripts[32,33]. SIA measures the correlation in intensity $g$ of two pixels in an image as a function of separation distance $r$. In practice, we generated autocorrelation curves $g(r)$ by taking the fast Fourier transform of the image $F(I)$, multiplying by its complex conjugate, applying an inverse Fourier transform $F^{-1}$, and normalizing by the squared intensity: $g(r) = \frac{F^{-1}(|F(I(r))|^2)}{[I(r)]^2}$. The data shown in Fig 5 is the average and SEM of $g(r)$ computed for all 10-20 images for a given composite.

## Acknowledgements

RMRA acknowledges funding from Air Force Office of Scientific Research, grant no. FA9550-21-1-0361. PMI acknowledges funding from the National Science Foundation, grant no. 2002362. We thank Ryan McGorty (University of San Diego) for guidance on rheology measurements.

## Author Contributions

preparing reagents: JS, SJ
conducting experiments: SJ, IML
analyzing data: SJ, GR, RMRA
supervising research: RMRA, PI
writing manuscript: RMRA, PI
editing manuscript: SJ, IML

## Data Availability

All data presented in the figures will be available on Zenodo when the manuscript is published.

## Declaration Statement

The authors declare no conflicting interests.

**Starch granules are instructive scaffolds for synergistic reinforcement and dissipation in hydrogel composites**

Shirlaine Juliano[1], Jasmine Samaniego[2], Ian M Lillie[1], Geraldine Ramirez[1], Peter M Iovine[2*], Rae M Robertson-Anderson[1*]

[1]Department of Physics and Biophysics, University of San Diego, San Diego, CA 92110
[2]Department of Chemistry and Biochemistry, University of San Diego, San Diego, CA 92110

*randerson@sandiego.edu, piovine@sandiego.edu

**Supplemental Information**

**Figure S1.** Linear oscillatory rheology measurements with a range of strain amplitudes to identify the linear regime.

**Figure S2.** Brightfield microscopy images of starch granules in single-type starch-gelatin composites.

**Figure S3.** Brightfield microscopy images of starch granules in blended starch-gelatin composites.

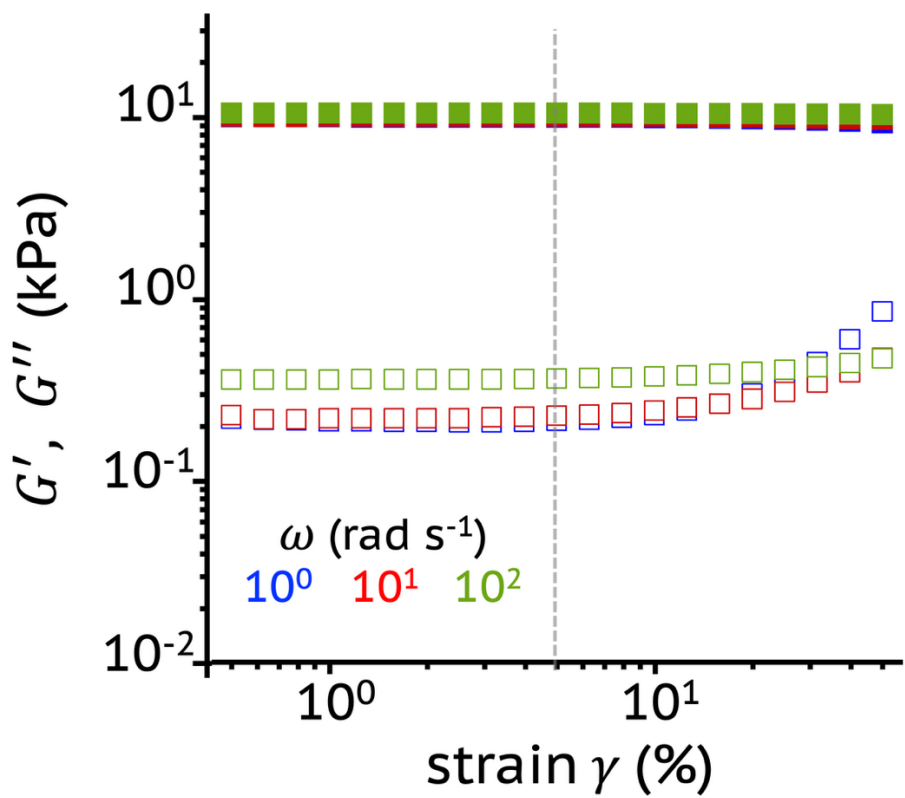

**Figure S1. Linear oscillatory rheology measurements with a range of strain amplitudes to identify the linear regime.** Linear viscoelastic moduli, $G'$ (filled symbols) and $G''$ (open symbols) as a function of oscillation strain amplitude $\gamma$ (%) for pure gelatin ($[s] = 0$). Strain sweeps were performed at oscillation frequencies of $\omega = 10^0$ (blue), $10^1$ (red), and $10^2$ (green) rad s$^{-1}$. Dashed vertical line denotes $\gamma = 5\%$, the strain amplitude at which all frequency sweeps presented in the main text are conducted.

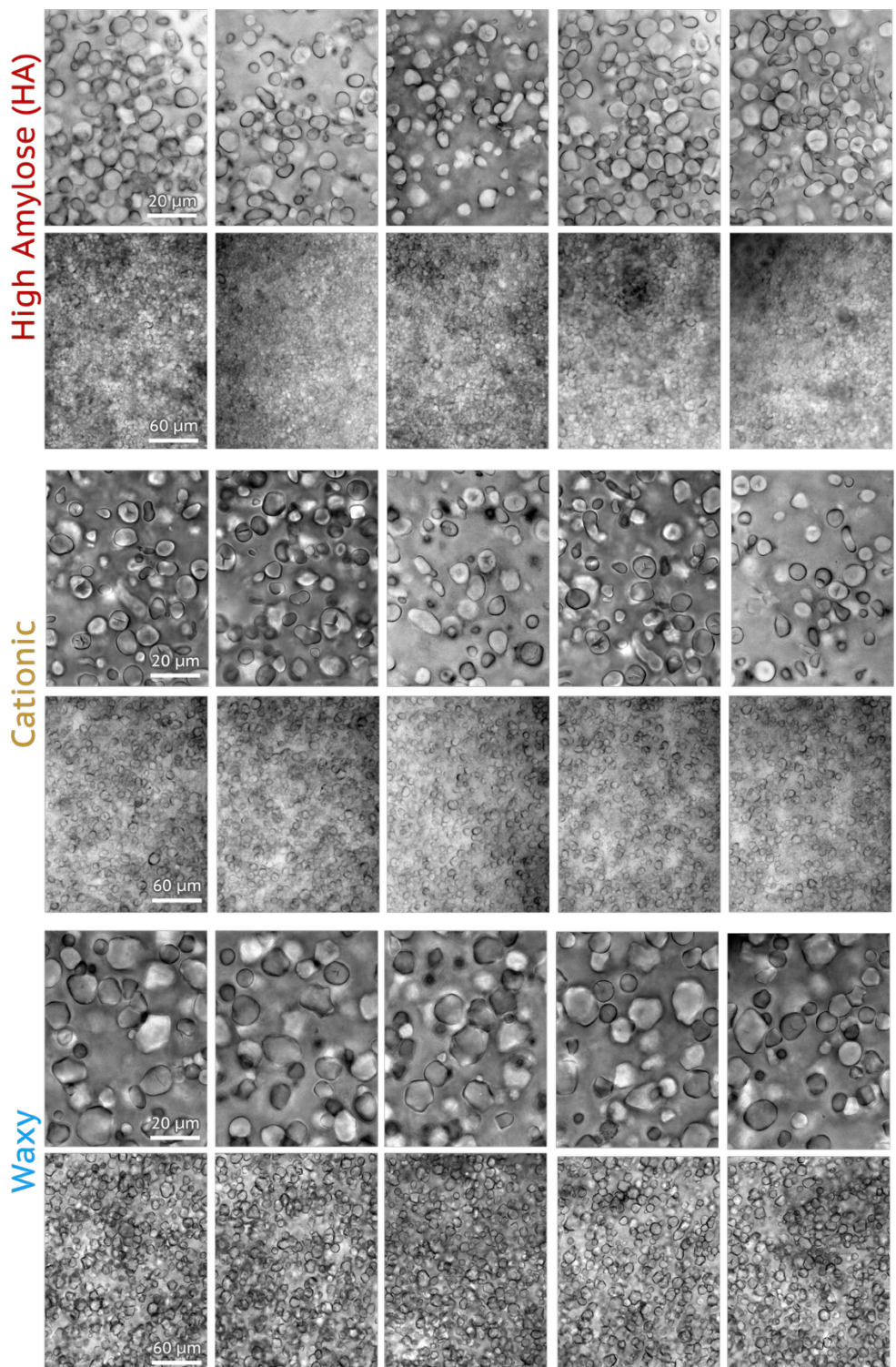

**Figure S2. Brightfield microscopy images of starch granules in single-type starch-gelatin composites.** Images collected at high (rows 1,3,5) and low (rows 2,4,6) magnification for composites comprising 16 wt% of HA (rows 1,2), Cationic (rows 3,4) or Waxy (rows 5,6) starches. Images in each row were collected in different regions of two samples. Scale bars in each row apply to all images in that row. See Methods section in main text for microscopy and acquisition details.

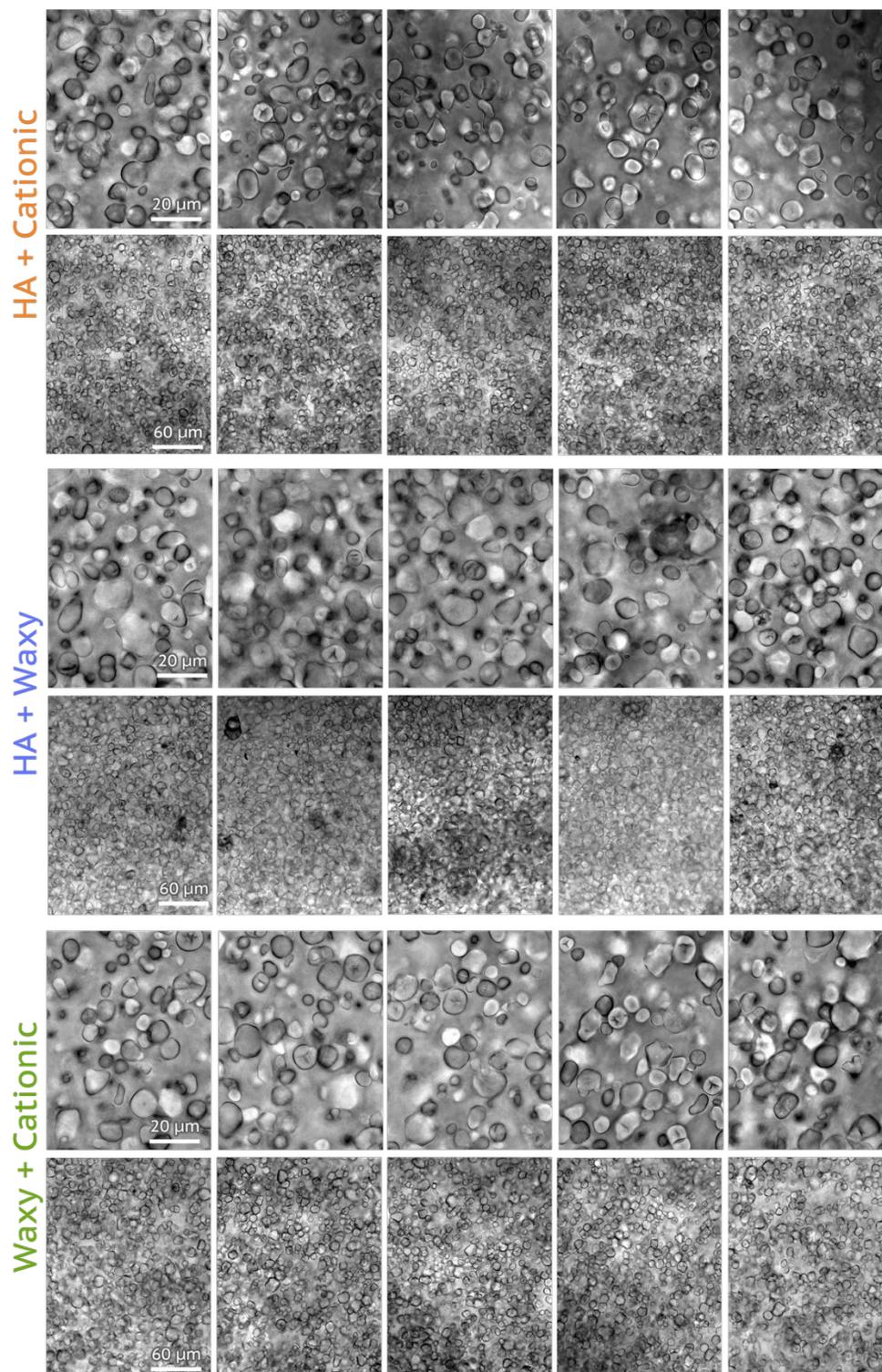

**Figure S3. Brightfield microscopy images of starch granules in blended starch-gelatin composites.** Images collected at high (rows 1,3,5) and low (rows 2,4,6) magnification for composites comprising blends of 8 wt% each (16 wt% total) of HA+Cationic (rows 1,2), HA+Waxy (rows 3,4) or Waxy+Cationic (rows 5,6) starches. Images in each row were collected in different regions of two samples. Scale bars in each row apply to all images in that row. See Methods section in main text for microscopy and acquisition details.

4